\documentstyle[11pt,aaspp4,epsfig]{article}

\newcommand{\boxsize}{0.98\textwidth}

\def\beq{\begin{equation}}
\def\enq{\end{equation}}
\def\bea{\begin{eqnarray}}
\def\ena{\end{eqnarray}}
\def\bec{\begin{center}}
\def\enc{\end{center}}
\def\etal{{\it et al.}}

\def\los{L.O.S.}
\def\msun{M_\odot}
\def\Mesz{M\'esz\'aros~}
\def\Pacz{Paczy\'nski~}
\def\eps{\epsilon}

\def\xfe{x_{Fe}}
\def\siml{\lower4pt \hbox{$\buildrel < \over \sim$}}
\def\simg{\lower4pt \hbox{$\buildrel > \over \sim$}}
\def\cc{\hbox{cm}^{-3}~}
\def\csq{\hbox{cm}^{-2}~}
\def\kal{K-$\alpha$~}
\def\ked{K-edge~}

\begin{document}
\slugcomment{ApJ subm. 8/21/99}

\title{ Early X-ray/UV Line Signatures of \\
  GRB Progenitors and Hypernovae }

\author{C. Weth$^{1,2}$, P. \Mesz$^{1,3,4}$, T. Kallman$^5$ \& M.J. Rees$^6$}

\noindent
$^1$Dpt. of Astronomy \& Astrophysics, Pennsylvania State University,
University Park, PA 16803 \\
$^2$Inst. f. Astron. \& Astrophysik, Univ. T\"ubingen, a.d. Morgenstelle 10, 72076 T\"ubingen, Germany\\
$^3$Institute for Theoretical Physics, University of California, Santa Barbara, CA 93106-4030\\ 
$^4$Astronomy Dpt., California Institute of Technology, MS 105-24, Pasadena, CA 91125 \\
$^5$Code 662, Lab. for High Energy Astrophysics, NASA/Goddard, Greenbelt, MD 20771\\
$^6$Institute of Astronomy, University of Cambridge, Madingley Road, Cambridge CB3 0HA, U.K. 

 

\begin{abstract}
We calculate the X-ray/UV spectral line signatures expected from the 
interaction of a gamma-ray burst afterglow and a dense pre-burst environment
produced by the progenitor. We explore the conditions under which Fe line
and edge equivalent widths of $\sim$ 1 keV can arise, and discuss the
possibility of gaining information about possible progenitor scenarios
using X-ray metal line spectra in the first few days of a burst.
A wind or supernova shell around the burst produces an X-ray absorption line spectrum 
and later emission lines, while a hypernova funnel model produces mainly emission lines. 
The Fe \ked can in some cases be more prominent than the Fe \kal line.
Under simple assumptions for the input continuum  luminosity, current reports of 
observed Fe line luminosities are compatible with an Fe-enriched funnel model, 
while lower values are expected in shell models.

\end{abstract}



\section{Introduction}

The nature of the progenitors of gamma-ray bursts (GRB) is an unsettled issue
of extreme interest, e.g. \cite{fryer99,pac98,mesz98}. It is becoming increasingly
apparent that whatever the progenitor, a black hole plus debris torus may result 
which powers the GRB, but the burning question is what gives rise to this system.
Both compact binary (NS-NS or BH-NS) mergers, other mergers (WD-BH, He-BH) or
the collapse of a massive, fast rotating star (referred to as hypernovae or 
collapsars) could lead to such a BH plus debris torus energy source, 
and much current work centers on discriminating between the various progenitors. 

Evidence concerning the progenitor comes both from the accumulating statistics on 
the off-sets between GRB afterglow optical transients and their host galaxies 
(\cite{bloom98}) and from light curve fits and continuum spectral information
providing evidence for either low (\cite{wiga99}) or high (\cite{owen98}) density 
in front of the afterglow. However, the most direct diagnostics  for the
environment are probably X-ray and UV spectral lines (\cite{bkt97,pl98,mr98b,bot98})
and an interesting possible diagnostic for hypernovae or collapsars is the
presence of Fe \kal emission lines, produced by fluorescent scattering from the
outer parts of the stellar progenitor of the continuum X-ray photons originating in
the afterglow of the GRB (\cite{mr98b,ghisel+99,lazz+99}). 

Quantitative calculations of spectral diagnostics of GRB progenitors are 
hindered by the lack of detailed  calculations or data on the evolution and
mass loss history in the period of months  to years before the outburst.
However, it is possible to guess what some of the generic features of such
pre-burst environments may be. 
The purpose of this paper is to consider some very simplified but physically
plausible progenitor configurations, and to explore in quantitative
detail the range of possible X-ray/UV spectral signatures that can be expected 
from them in the time scale of hours to days after the outburst.

\section{Pre-burst Environments and Computational Method}

The task of finding useful progenitor diagnostics is simplified if the pre-burst 
evolution of the latter leads to a significantly enhanced gas density in the immediate 
neighborhood of the burst. 
In the case of a massive progenitor scenario, 
such as a hypernova or collapsar, it is known that red supergiants and supernova progenitors 
in general are prone to have strong winds. One would expect such a strong mass loss phase to 
produce a pre-burst environment which could have the form of a shell, e.g. as inferred in 
SN 1987a. For instance, a star evolving from a red giant to a blue giant phase might emit first 
a slower wind, which is later swept up into a shell by a faster wind.  In some collapse or
compact binary merger scenarios, e.g. of a BH or NS with a White Dwarf or He core left over 
from a massive companion star, a supernova producing a metal-enriched supernova remnant (SNR) 
shell might precede the burst. In general the shell would be expected to have dispersed before 
the burst occurs, but there could be rare cases where this is not the case. Another possible
scenario which has been discussed is the delayed collapse of a rotationally stabilized neutron 
star, which could lead to a burst with a SNR shell around it (\cite{VietriStella98}).

Another geometry characterizing massive progenitor or collapsar models could arise
if the giant progenitor is fast-rotating, e.g. due to spin-up from
merging with a compact companion. Then the stellar envelope and the wind would
be expected to be least dense inside a funnel-like cavity extending along the
orbital-spin axis, with the GRB at the tip of the funnel.
However, detailed models for either a funnel-like environment or for a shell resulting
from a GRB progenitor are lacking so far. Therefore, our choice of parameters below for
these scenarios is purely phenomenological, and guided more by the reported observations than
by theoretical considerations.

For the computations, we need to treat in detail the photoionization and
recombination of the various ions in the environment material, and to
obtain spectra which can be compared to observations we need to consider
the time-dependence. The latter is due to the fact that the recombination and 
ionization has a natural timescale depending on the ambient density, the 
chemical abundance and the flux received; that the ionizing spectrum from the 
GRB afterglow varies in time;  and that the spectrum observed at a given observer 
time is made up of light arriving from different regions of the remnant, for 
which the source time is different. The problem can be simplified if the first 
timescale is shorter than the latter two, since in this case one may use a 
steady-state photoionization code.  
The recombination time is $t_{rec}\sim 10^3 Z^{-2} T_7^{1/2} n_{10}^{-1}$ s for ions
of charge $Z$ at the typical temperatures and densities in the reprocessing gas, which
is short compared to the timescales $\sim 10^5 {\rm s} \sim$ 1 day considered, so the 
ionization equilibrium approximation is justified in the examples calculated below.
In this paper we exploit this approximation, 
and make use of the XSTAR code (\cite{kallman82,kallman98}) to calculate the spectrum. 
This is a steady state code which, for a given input spectrum, 
calculates the photoionization of a plasma in a shell at a given 
distance from the source, as a function of the density and chemical abundances.
These position dependent spectra in the source frame, which arise in response to 
a time-variable input spectrum, are integrated over the remnant to obtain the 
observer-time dependent spectra that would be actually measured. 
A restriction on the use of this code is that the effects of comptonization
can be included only in a rough manner. This is not a problem if the remnant
is Thomson thin (column density $\Sigma \siml 2.5 \times 10^{24}\csq$), or if 
the incident continuum is absorbed over a column density smaller than this.
Very little if at all is known about the nature and geometry of the remnants,
and in what follows we assume situations where the above restriction is either 
satisfied, or the effects of its violation can be estimated by means of a 
different Monte Carlo code (\cite{matt96}) which is not subject to this restriction.

The input spectrum that we assume is typical of simple afterglow models (\cite{mr97a,wax97}), 
with phenomenological parameters chosen to approximate those of the
observed afterglow GRB 970508. We take these to be a break luminosity $L_{\eps_m} 
\simeq 3.2\times 10^{46}$ erg/s/keV with a break energy $\eps_m=1.96$ keV at 
$t=10^3$ s, a \cite{band+93} spectrum with energy indices $\alpha=0.33$, $\beta=-0.75$ and a 
standard time decay exponent for the peak frequency of $\gamma= -(3/2)\beta$. 

\section{ Shell Models  }

One type of environment model considered is a shell of gas at some distance from the burst, 
considered to be essentially stationary over the period of interest for its response to the above 
time-dependent input spectrum. The shells could be metal-enriched, especially if arising from a 
supernova explosion before the burst, and possibly also if involving a pre-ejected wind from 
a massive stellar progenitor (although in the latter, solar abundances are probably likelier). 
These shells could have a large coverage fraction, and would have a mean density much 
larger than typical ISM values, especially when blobs and condensations form via instabilities.  
Guided by the report of an Fe line detection peaking after about one day in GRB 970508 and in 
GRB 970828 (\cite{PiroFe99,yoshida98}), one is led to consider shell radii $R \simg$ light day. 
A physical requirement is that the distance $ct\Gamma(t)^2$ reached after one day by the 
afterglow shock producing the continuum be less than $R$, with $\Gamma\simg 1$.
As an example, we assume that the shock is observed to reach the shell at an inner-radius 
distance $R\sim 1.5\times 10^{16}$ cm at $t_s=1$ day along the \los.
Within the context of simple (adiabatic, impulsive, homogeneous external density) standard 
afterglow models, this could occur for a deceleration radius $\siml R$, requiring a 
density between the burst and the shell higher than usually considered. A pre-shell density 
$n \siml 10^6 \cc$ could do this, involving a total mass $\sim 10^{-2}\msun$ much less than in the 
assumed shell, and a Thomson optical depth $\tau_T \ll 1$.  However, in scenarios leading to 
a shell the conditions might differ substantially from those implied in snapshot fits 
to simple standard models, e.g. Wijers \& Galama 1999, and the error bars in such fits are hard 
to estimate. A complete model of the physics for both the input continuum and the reprocessing gas 
would be uncertain, especially in view of the preliminary nature of current X-ray line observations.
For this reason, we prefer to consider a phenomenological input spectrum as a quantity
given by observations, and treat the environment simply as a test particle gas, choosing its 
physical parameters in such a manner as to reproduce the current observations. 

For computational reasons, the calculations are carried out for thin 
homogeneous spherical shells of different densities, which can be used to 
represent thicker inhomogeneous shells with the same mass per unit area and the 
same density in the filaments or blobs as in a homogeneous thin shell.
The shell was assumed to have an Fe abundance either 10 times solar or 
$10^2$ solar (and solar for the other elements), and a hydrogen column density
$N_H= 5\times 10^{23}\csq$, with a total mass $M_s\sim 1\msun$. 
The X-ray/UV spectrum as a function of observer time is shown in Figure 
\ref{fig:sh5sp}, for several values of the particle density in the shell.
%
%
The line spectrum becomes more prominent as the gas cools and recombines. Due to the 
very high luminosities and hard initial $\gamma$-ray spectrum, initially all the Fe 
is fully ionized, and as it cools the strongest features initially 
are the Fe \kal and \ked, which initially appear in absorption and later as
recombination {\it emission} features, the strength of the \ked feature peaking at
later times than the \kal for these luminosities, the \kal line being more prominent and
easier to detect for this input continuum luminosity. (For a lower input continuum 
luminosity with a steeper spectrum, however, the time sequence can reverse and the 
\ked can be more prominent than the \kal feature). As the continuum continues to
decrease the Fe \kal line becomes more important, shifting its energy gradually 
from 6.7 to 6.4 keV as the lower ions become in turn more predominant. 
At later times the Fe \ked recombination emission feature begins to emerge in emission
as well, whereas at early times times it is largely an absorption feature.
After the Fe features have become important, with some delay 
depending on the density and abundance, other features in the 2-3 keV range due 
to Si and S also become prominent, as well as an O recombination and \kal features 
at 0.86 and 0.65 keV. 

%
%
The corresponding X-ray light curves in the 2-10 keV range are shown in Figure 
\ref{fig:sh5lc}, as well as the equivalent widths of the Fe \kal feature. 
The Fe \kal luminosity reaches values $\siml 10^{43}$ erg s$^{-1}$ and the EW reaches values 
of 0.2 to 3.5 keV in emission.
The Fe \ked feature in emission reaches values $\siml 0.1$ keV in Figure 2d (not shown),
but at early times the \ked absorption is substantial, as seen in Figure \ref{fig:sh5sp}.
In this example of a full shell the equivalent width (EW) of the Fe \kal in the 
6.4-6.7 keV range and the Fe \ked at $\sim 9.28$ keV continues to grow as the bulk of the 
diffuse \ked recombination and fluorescent \kal photons reach the observer from 
the rim and the back portions of the shell, in response to the GRB time-dependent 
continuum. This growth continues until a time $t\sim R/c \sim 5$ days, when the 
diffuse radiation from the rim of the shell becomes visible.
However, by this time the total X-ray flux (continuum plus lines) 
has decreased significantly (Figure \ref{fig:sh5sp}) and the S/N is less 
favorable for detection. 
%
%
Notice that in this calculation the continuum source, i.e. the shock, crosses the
shell at 1 day. At this point the observed continuum X-ray luminosity temporarily 
increases, as the radiation along the \los is no longer absorbed by the shell,
but then it continues to decrease according to the standard afterglow decay law. 
This temporary brightening would be enhanced, 
and might be dominated by, the heating of the shell as the shock goes across it; 
a consistent analysis of the shock heating would require a number of additional assumptions and 
detailed gas dynamical calculations which are beyond the scope of this paper (see, e.g., 
\cite{vietri+99} for an analytical estimate). 
A temporary brightening of the continuum at one day is in fact seen in the 
observations of GRB 970508 (\cite{PiroFe99}).
The unabsorbed continuum reaching the
observer after one day from beyond the shell is also responsible for the gradual 
re-filling of the absorption throughs seen in Figure \ref{fig:sh5sp} at late times.
%
%

The effects of a jet-like fireball illuminating a spherical shell is  also of
interest. An example of the spectral evolution is shown in Figure \ref{fig:bm5sp}, 
for a fireball whose continuum radiation is collimated in a jet of opening 
half-angle $\theta_j=37\deg$ (and other properties the same as for the spherical 
fireball of Figure \ref{fig:sh5sp}). In this example the shell was assumed to be 
spherical, with the same dimensions and properties as in Figure \ref{fig:sh5sp}.
The effect of a jet is that the ring-shaped area of illuminated shell which is 
visible to the observer increases only up to a time $t_j=(R/c)(1-\cos\theta_j) 
\sim$ 1 day. After that time, the shell regions at angles larger than $\theta_j$ 
which become visible do not contribute any diffuse radiation, since they are not 
(and were never) illuminated by the continuum source. This choice of $\theta_j$ 
results therefore in \ked and \kal equivalent widths which grow until $t_j\sim 1$ 
day, and decay thereafter (see Figure \ref{fig:bm5lc}).

\section{Scattering Funnel Hypernova Models }

A different configuration which may characterize hypernovae involves a funnel geometry.
Accurate hypernova line diagnostics will be uncertain due to the absence of quantitative 
models, extending from minutes to days after the burst, of the gaseous environment
in the outer layers and/or winds in such objects. We can, however, get an estimate
of what may be expected by using a physically plausible toy model.  We take a 
parabolical funnel as an idealized representation of the centrifugally 
evacuated funnel along the rotation axis of the collapsing stellar configuration, 
with the GRB at its tip. In order to produce line features which peak at about
one day from such a model, one requires the X-ray continuum to be inside 
the outer rim of the funnel for at least this long. A simple configuration 
with these properties is, for example, a wind with a scattering optical depth 
$\simg 1$ extending out to $R=1.5\times 10^{16}$, in which there are two empty (or 
at any rate much lower density) funnels, inside which which the fireball expands. 
%
%
The fireball is assumed to have the same luminosity per solid angle and
spectral characteristics as used in the previous two shell models, and
the funnel opening half-angle was taken to be $15\deg$. 
For the funnel walls we take a uniform density $n=10^{10}\cc$, and an 
Fe abundance $\xfe=10$ or $\xfe=10^2$; 
we assume the effective column density within which 
reprocessing is most effective to be $\Sigma=10^{24}\csq$, the effective
amount of reprocessing mass involved being $\sim 0.2\msun$.
An accurate calculation of the spectrum 
escaping from a funnel is not straightforward, since a rigorous prescription for 
treating multiple scatterings and a non-spherical geometry is difficult to
implement in a code such as XSTAR.  However, it is possible 
to obtain useful lower and upper limits for the actual equivalent widths, by 
calculating the widths expected in two limits. A low estimate for the EW is
computed by counting only the once-reflected line photons which are directed
inside the opening angle of the funnel, and comparing them to the continuum
photons (either direct or reflected) which are similarly directed inside the 
opening angle. The high limit for the EW is calculated using all the
once-reflected line photons (whether directed at the opening or not) and
comparing them to the directly escaping plus all the reflected continuum.
A spectrum as a function of time for the second limit (all) and $\xfe=10^2$ is shown in
Figure \ref{fig:hnn2}. 
%
%

The funnel model was taken to have the same input luminosity per solid angle as the shell 
models, but the incidence angle is shallower in funnels, and hence the effective heating per 
unit area is smaller than in shells at the same distance, which favors Fe \kal recombination, 
hence the Fe \kal luminosity is larger. The upper and lower limits for this hypernova example 
with $\xfe=10^2$ (see bottom of ~figure \ref{fig:hnn2}) show that at one day the Fe \kal 
luminosity is bounded between $2\times 10^{44}$ erg s$^{-1}$ and $6\times 10^{42}$ erg s$^{-1}$, 
and the Fe \kal line EW is bounded between 1.2 and $\sim 0.1$ keV, while the Fe \ked EW,
which is more prominent, is bounded between 2.7 and 0.2 keV. 
For $\xfe=10$, these values are lower by a factor $\sim 3$ (Figure \ref{fig:hnn1}).

\section{Discussion}

We have considered a series of models where the environment of the burst can be 
represented as a shell of enhanced density at some radial distance from the burst. 
These shells could be the result of a pre-burst wind phase of a massive  progenitor, 
a hypernova, collapsar or a merger involving a massive companion or its core. 
Alternatively, they might be supernova remnant (SNR)  shells of a rare kind, which 
originated sufficiently recently that they have not yet dispersed before the burst occurs.
Such shells can produce significant Fe \kal and \ked luminosities and equivalent widths of
order $\sim$ keV, provided the density (possibly in the form of blobs) in the shell is 
large ($10^{10} - 10^{12} \cc$), and the coverage fraction is a substantial fraction of $4\pi$.
For a mass of Fe in the shell $\sim 2.5\times 10^{-4}\msun$ or $2.5\times 10^{-3} \msun$ 
(a total shell mass $1\msun$) the Fe \kal equivalent widths after one day can be
$E_W \siml$ keV, comparable to the values reported by \cite{PiroFe99} and \cite{yoshida98}. 
However, the Fe luminosity is $\siml 10^{43}$ erg s$^{-1}$, which is low by a factor of 5-10.
The higher density or more Fe-rich shells also show a drop in the continuum after 
$t \sim 10^4$ s due to Fe absorption and re-emission, e.g. panels b.c and d of Figure 2,
a feature which qualitatively resembles an observed dip in the light curve of GRB 970508
at $\sim 5\time 10^4$ s (\cite{PiroFe99}). 
We note that winds of ${\dot M} \sim 10^{-4} \msun$/yr with velocities $v_w \sim 100$ Km/s 
varying on timescales $\siml 100$ years (characteristic of massive stars) would yield shell 
enhancements starting at $R\sim 10^{16}$ cm with mean density $n\simg 10^5\cc$, 
in which condensations of $n\sim 10^9-10^{11}\cc$ could form via instabilities. 
Dense shells may also form as a result of a fast wind following a slower one.
The results would be similar whether the shells are homogeneous, or consist of 
blobs with the same density and a comparable total coverage fraction as a homogeneous shell. 
(Note that in this model the shell or blobs have a different origin, are further out and are 
much bulkier and slower than, e.g the metal-enriched blobs possibly accelerated in the 
relativistically moving burst ejecta itself, e.g. \cite{hailey99,mr98a}).

In this particular model, the Fe \kal EWs reach values $E_W \sim 0.3-3$ keV at 1 day, 
and continue to be significant up to a time $\sim R/c\sim 5$ days when the diffuse radiation 
from the rim of the shell reaches the observer (Figures \ref{fig:sh5sp} and \ref{fig:sh5lc}). 
However the continued decay of the continuum after 1 day would reduce the S/N ratio, which 
would make it harder to detect an Fe feature at later times.

We have also explored a different shell scenario, where the Fe features would 
cut-off abruptly after reaching a peak. This occurs if the continuum is beamed,
e.g. it is produced by a collimated fireball jet. In this case (Figures 
\ref{fig:bm5sp} and \ref{fig:bm5lc}), the diffuse radiation, including the Fe and
other spectral features, cuts off after a time $t\sim (R/c)(1-\cos\theta_j)
\sim 1$ day, and for a $\sim 1\msun$ shell at $R\sim 10^{16}$ cm with
Fe abundance $10-10^2$ times solar the models produce Fe \kal equivalent
widths $\siml 0.3-3$ keV (depending on the density) peaking at one day.

Another series of models that we considered address the consequences of a
funnel geometry in a spinning massive progenitor (hypernova or collapsar). 
If the stellar envelope or its wind can be assumed to extend out to radii of
light-days with an appreciable density of order $n\sim 10^9-10^{11}\cc$ 
(particularly in winds where the density drops slower than $r^{-2}$, e.g. as 
in hour-glass shaped winds such as that of SN 1987a), 
a wider range of luminosities and equivalent widths is possible after $\sim 1$ day. 
In this case, for $\xfe=10^2$ we obtain an Fe \kal luminosity $L_{Fe}\sim 2\times 10^{44}$ 
erg s$^{-1}$ at $t\sim 1$ day (or a factor 3 lower, for $\xfe=10$), comparable to the 
observational results.
The equivalent width grows until the continuum source (or afterglow shock) moves beyond
the radius where there is a substantial amount of stellar wind material to
reprocess it (see Figure \ref{fig:hnn2}). After that time, only the decaying
shock continuum is detected which is now beyond the wind region, and a fast 
decaying component from the funnel wall as it cools.
 
The calculations presented here indicate that a qualitative difference between
shell and funnel models is that, whereas shells produce Fe \kal, \ked and features
from other metals predominantly in absorption, and later also partly in emission,
the funnel models are dominated by emission features throughout. 
This is due to the presence of material along the line of sight in the shell 
models, which is absent in the funnel case.  

It is worth noting that in GRB 970508 the energy of the X-ray spectral feature discussed by 
\cite{PiroFe99} agrees with that of a 6.7 keV Fe \kal line at the previously known redshift 
$z=0.835$ (\cite{metz+97}), while in GRB 970828 the energy of the X-ray spectral feature 
reported by \cite{yoshida98} is compatible with an Fe \ked feature at 9.28 keV in the rest 
frame, at the recently reported redshift $z=0.958$ (\cite{djorg+00}).

A general point is that in the case of low mass binary mergers, such as NS-NS or 
BH-NS, it is harder to see how shells or funnels would have formed and still be 
present within distances $\simg 10^{15}-10^{16}$ cm at the time of the burst.
Hence the detection of Fe \ked and \kal features peaking at $\sim$ 1 day at the 
strengths discussed here (and as reported by \cite{PiroFe99,yoshida98}) 
would appear to be a significant diagnostic for a massive progenitor. Shells and
funnels with dimensions about a light-day are rough examples of extreme geometries
which might characterize massive progenitor remnants. However, a clear distinction 
between various types of massive progenitors (or mergers involving a massive 
progenitor) would require extensive quantitative calculations in the spirit of, 
e.g. \cite{fryer99} and \cite{ruja99}, but considering 
more specifically the different pre-burst evolution and near-burst environments.
What our present calculations are able to indicate is that Fe \kal equivalent
widths of $\sim$ keV can be produced in a variety of plausible progenitor scenarios, 
but the absolute value of the Fe \kal line flux provides constraints on the combined values 
of the density, chemical abundance and distance from the burst, as well as the geometry.
Our present calculations, which include a number of simplifying assumptions, indicate that 
Fe-enriched funnel models agrees better than shell models with the currently reported Fe line 
values. More detailed modeling, as well as more sensitive X-ray spectral line detections, 
should be able to provide valuable constraints on specific progenitors. 

\acknowledgments
We are grateful to NASA NAG5-2857, NSF PHY94-07194, the Division of Physics, Math \& Astronomy,
the Astronomy Visitor Program and Merle Kingsley fund at Caltech, the DAAD and the Royal Society 
for support. We thank M. B\"ottcher for pointing out a significant discrepancy, and 
N. Brandt, G. Chartas, G. Djorgovski, A. Fabian, S. Kulkarni, A. Panaitescu, S. Sigurdsson 
and A. Young for discussions.


 

\newpage
\setcounter{figure}{0}


\begin{figure}[ht]
\begin{center}
\epsfbox{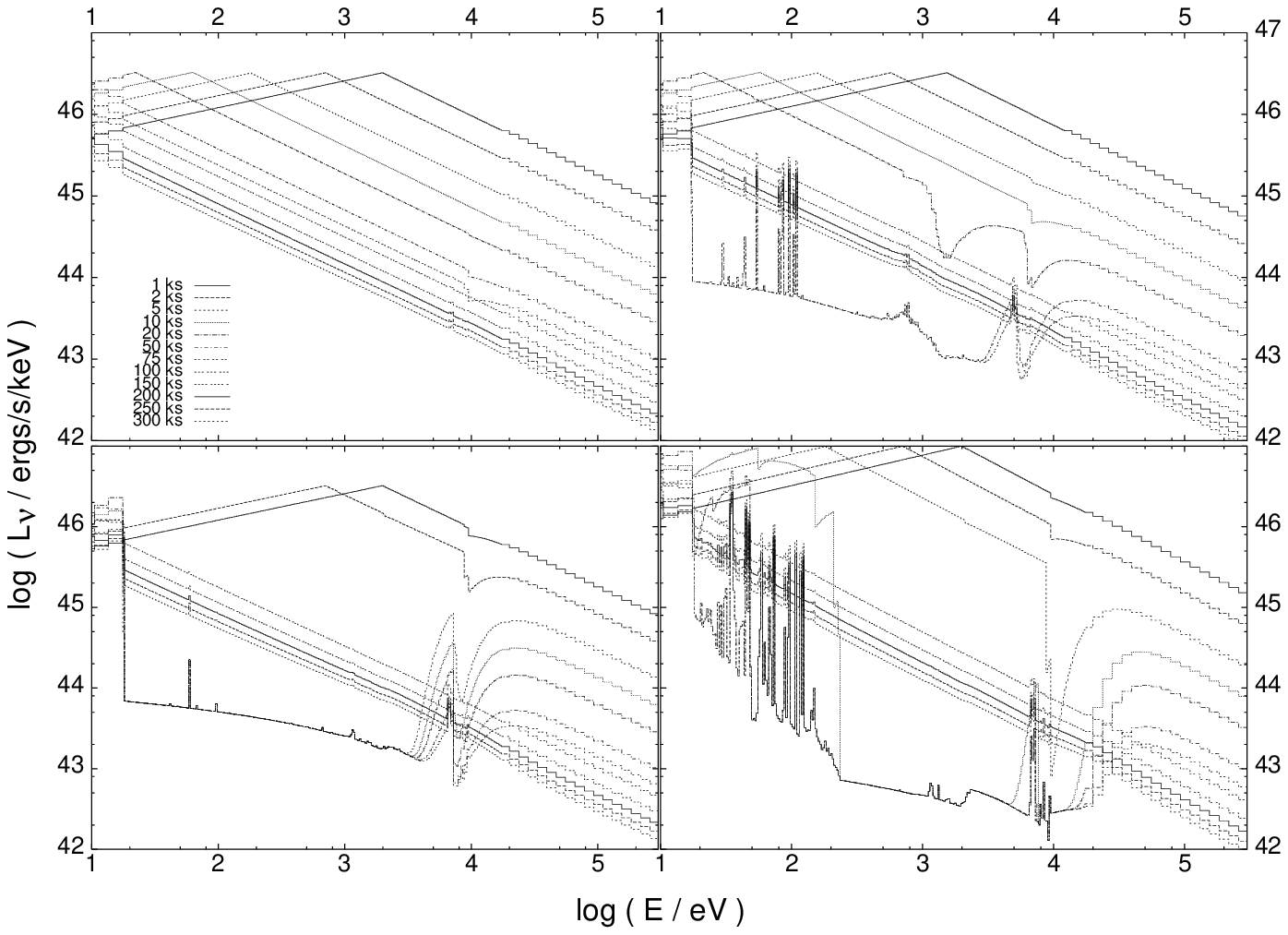}
\caption{\label{fig:sh5sp}
Spectrum of a shell model for various observer times (in seconds), with
$R=1.5\times 10^{16}$ cm, hydrogen column density $\Sigma= 5\times 10^{23}\csq$,
for several particle densities $n$ and Fe abundance $\xfe=n_{Fe}/n_{Fe,\odot}$.
Top left(a): $n=10^{10}\cc$, $\xfe=10$; top right(b): $n=10^{11}\cc$ $\xfe=10$;
bottom left(c): $n=10^{12}\cc$, $\xfe=10$; bottom right(d): $n=10^{11}$, $\xfe=10^2$.}
\end{center}
\end{figure}


\begin{figure}[ht]
\begin{center}
\epsfbox{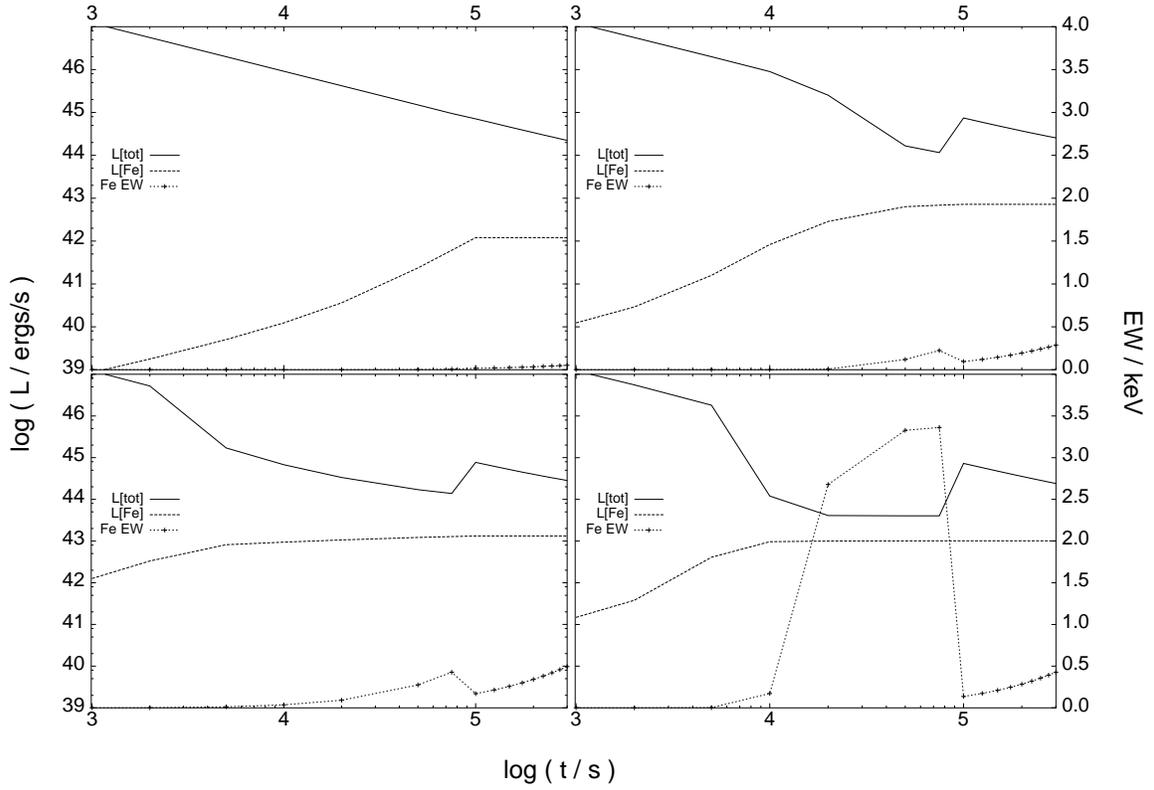}
\caption{\label{fig:sh5lc}
Light curves and equivalent width of a shell model,
same parameters as in Figure \ref{fig:sh5sp}. The $L_{tot}$ is the incident plus
reprocessed luminosity in the 2-10 keV range, and $L_{Fe}$ is the line luminosity
in the Fe \kal range.
Top left(a): $n=10^{10}\cc$, $\xfe=10$; top right(b): $n=10^{11}\cc$ $\xfe=10$;
bottom left(c): $n=10^{12}\cc$, $\xfe=10$; bottom right(d): $n=10^{11}$, $\xfe=10^2$.}
\end{center}
\end{figure}


\begin{figure}[ht]
\begin{center}
\epsfbox{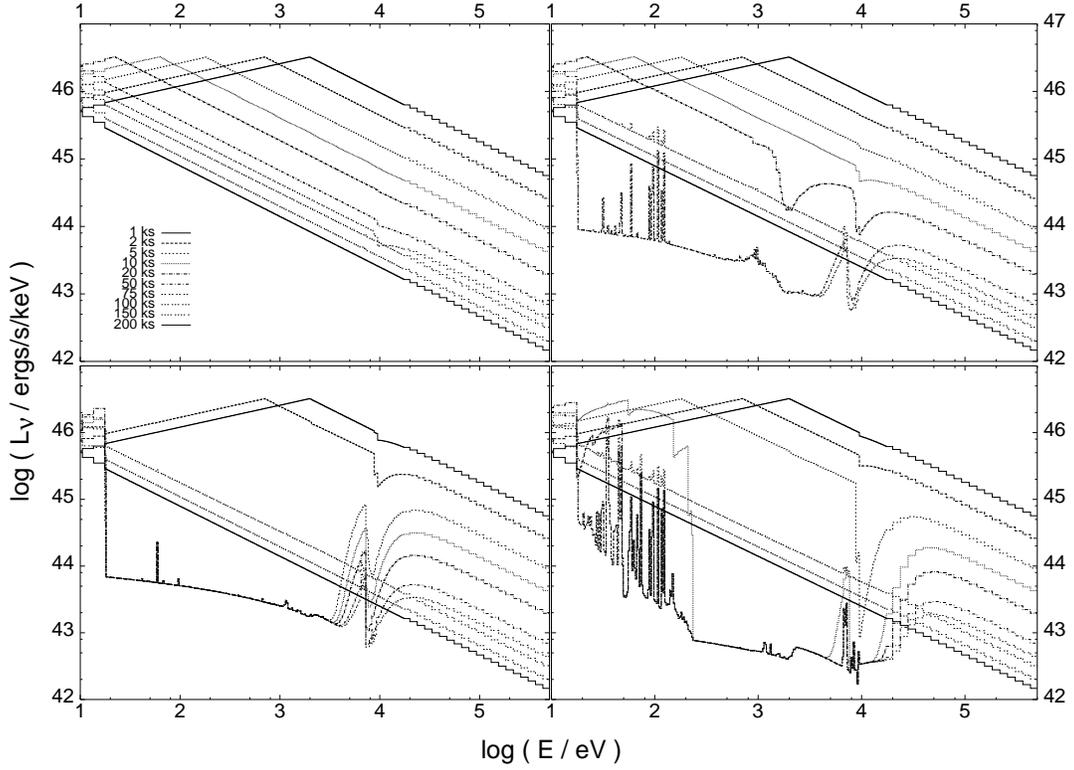}
\caption{\label{fig:bm5sp}
Spectrum of a shell model illuminated by a jet-like fireball ($\theta_j=37\deg$)
for various observer times (in seconds), with shell parameters and other fireball
properties the same as in Figure \ref{fig:sh5sp}.
Top left(a): $n=10^{10}\cc$, $\xfe=10$; top right(b): $n=10^{11}\cc$ $\xfe=10$;
bottom left(c): $n=10^{12}\cc$, $\xfe=10$; bottom right(d): $n=10^{11}$, $\xfe=10^2$.}
\end{center}
\end{figure}


\begin{figure}[ht]
\begin{center}
\epsfbox{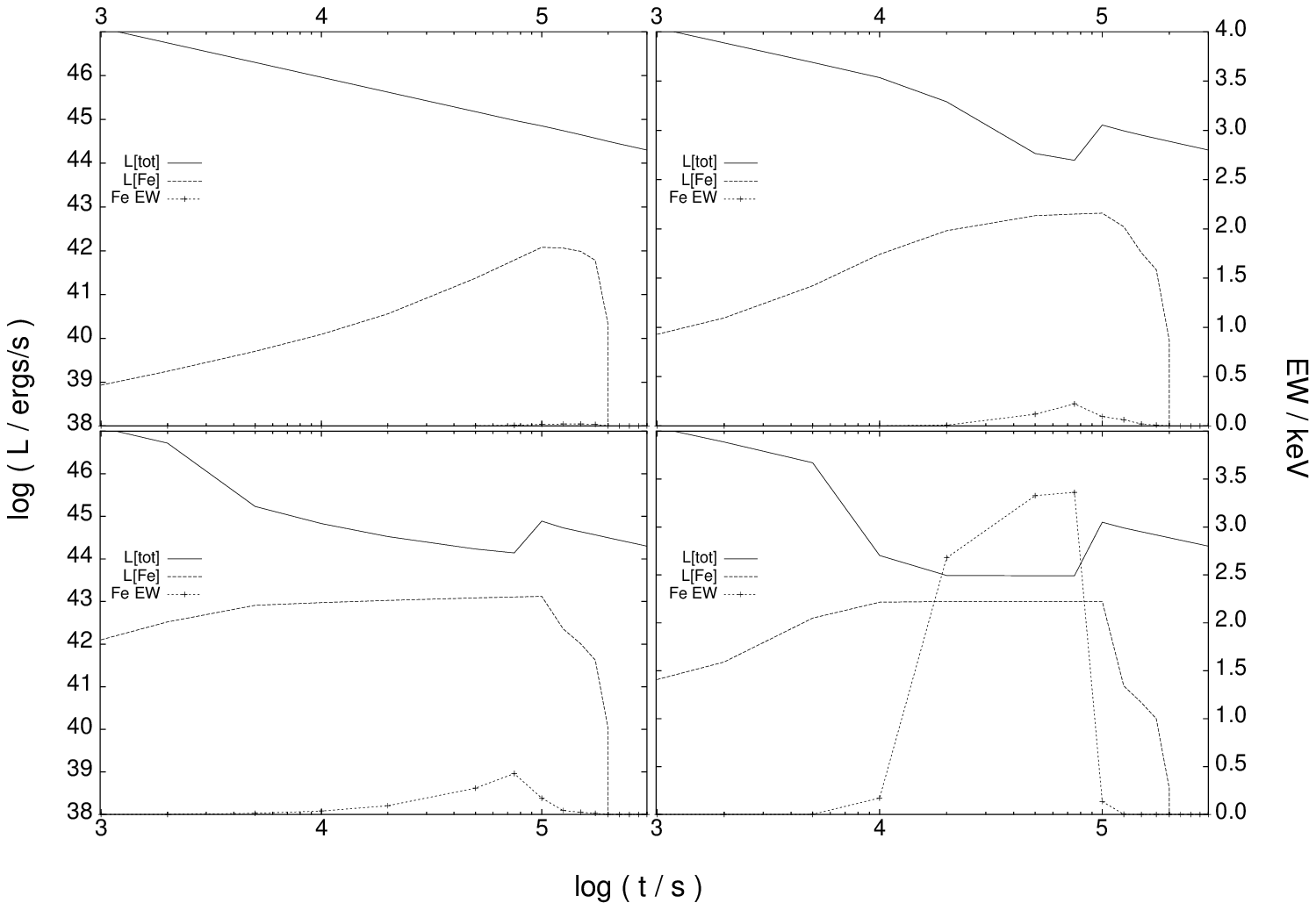}
\caption{\label{fig:bm5lc}
Light curves and equivalent width of a shell model
illuminated by a jet-like fireball,
same parameters as in Figure \ref{fig:bm5sp}. The $L_{tot}$ is the incident plus
reprocessed luminosity in the 2-10 keV range, and $L_{Fe}$ is the line luminosity
in the Fe \kal range.
Top left(a): $n=10^{10}\cc$, $\xfe=10$; top right(b): $n=10^{11}\cc$ $\xfe=10$;
bottom left(c): $n=10^{12}\cc$, $\xfe=10$; bottom right(d): $n=10^{11}$, $\xfe=10^2$.  }
\end{center}
\end{figure}

\vfill\eject


\begin{figure}[ht]
\begin{center}
\begin{minipage}[t]{0.5\textwidth}
\epsfxsize=\boxsize
\epsfbox{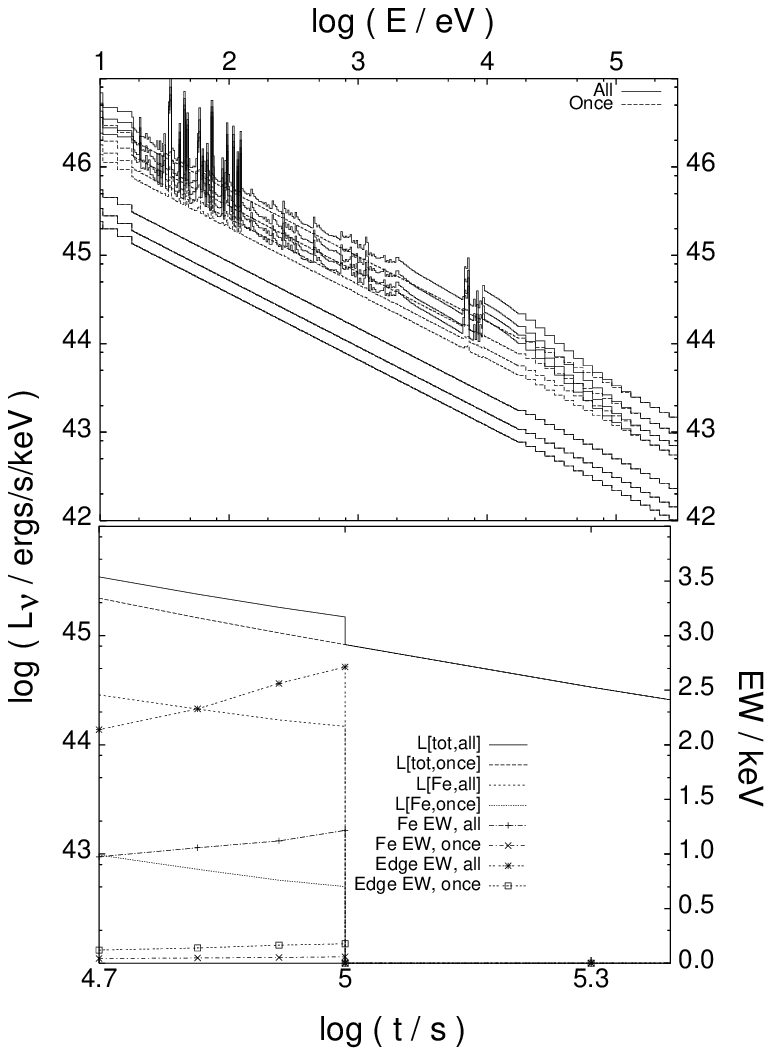}
\end{minipage}
\begin{minipage}[t]{\textwidth}
\caption{\label{fig:hnn2}
{\it Top}: Spectrum of scattering funnel model with $\xfe=10^2$ as a function of time (from
top to bottom 50, 66, 83, 100, 200 and 300 kseconds). The spectral luminosity is
shown in the upper/lower bound approximation (all/once, see text).
{\it Bottom}: Total and Fe light curves and Fe \ked and \kal equivalent widths for the
scattering funnel model as a function of time. The values are calculated for
the upper/lower bound approximations (all/once, see text).
     }
\end{minipage}
\end{center}
\end{figure}

\vfill\eject


\begin{figure}[ht]
\begin{center}
\begin{minipage}[t]{0.5\textwidth}
\epsfxsize=\boxsize
\epsfbox{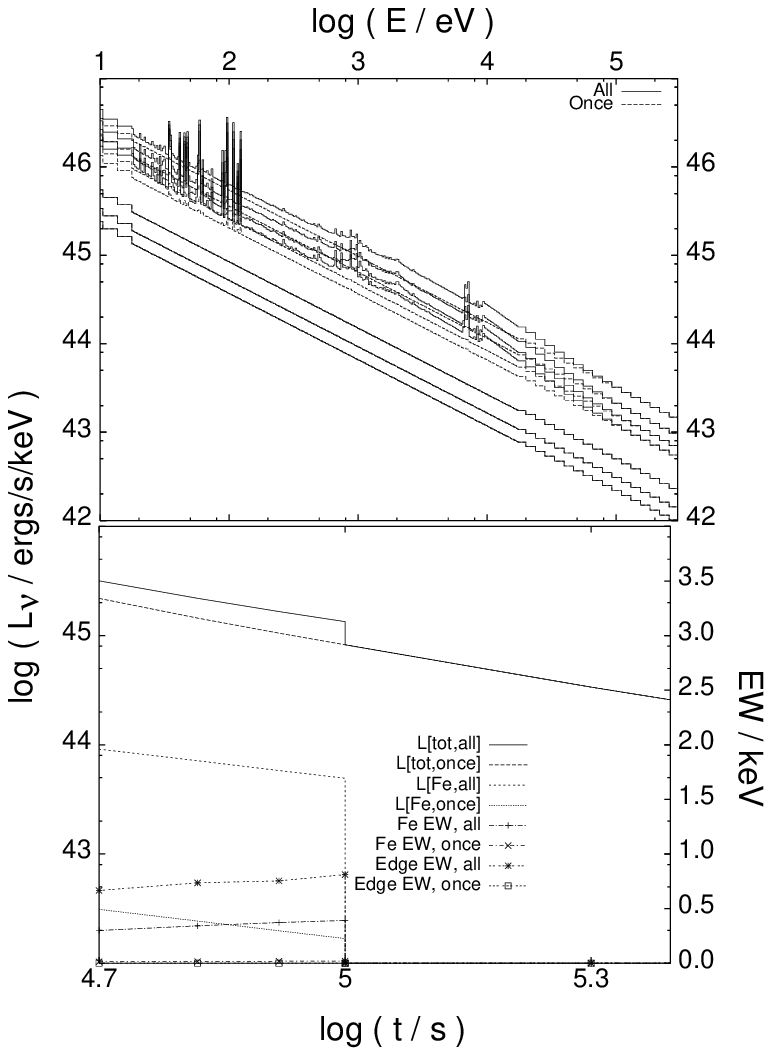}
\end{minipage}
\begin{minipage}[t]{\textwidth}
\caption{\label{fig:hnn1}
{\it Top}: Spectrum of scattering funnel model with $\xfe=10$ as a function of time (from
top to bottom 50, 66, 83, 100, 200 and 300 kseconds). The spectral luminosity is
shown in the upper/lower bound approximation (all/once, see text).
{\it Bottom}: Total and Fe light curves and Fe \ked and \kal equivalent widths for the
scattering funnel model as a function of time. The values are calculated for
the upper/lower bound approximations (all/once, see text).
     }
\end{minipage}
\end{center}
\end{figure}

\end{document}